\documentclass[journal]{IEEEtran}
\IEEEoverridecommandlockouts
\usepackage{cite}
\usepackage{amsmath,amssymb,amsfonts}
\usepackage{algorithmic}
\usepackage{times}
\usepackage{fixltx2e}
\usepackage{epsfig}
\usepackage{graphicx}
\usepackage{amsmath}
\usepackage{amssymb}
\usepackage{diagbox}
\usepackage{arydshln}
\usepackage{lscape}
\usepackage{caption}
\usepackage{subcaption}
\usepackage{colortbl}
\usepackage{multirow}
\usepackage{booktabs}
\usepackage{arydshln}
\usepackage{blindtext}
\usepackage{textcomp}
\usepackage{xcolor}
\usepackage{fancyhdr}
\usepackage{hyperref}
\def\BibTeX{{\rm B\kern-.05em{\sc i\kern-.025em b}\kern-.08em
    T\kern-.1667em\lower.7ex\hbox{E}\kern-.125emX}}
\hypersetup{
	colorlinks,
	linkcolor=[rgb]{1.0, 0.0, 0.0},
	citecolor=[rgb]{1.0, 0.2, 0.2},
	urlcolor =[rgb]{0.0, 0.0, 0.8}
}

\begin{document}

\title{PanNuke Dataset Extension, Insights and Baselines
}

\author{Jevgenij Gamper$^*$, Navid Alemi Koohbanani$^*$, Ksenija Benes, Simon Graham, Mostafa Jahanifar, Seyyed Ali Khurram, Ayesha Azam, Katherine Hewitt and Nasir Rajpoot
        \thanks{$^*$ First authors contributed equally.}
		\thanks{J.Gamper, N.A.Koohbanani, S.Graham and N.Rajpoot are with the Department of Computer Science, University of Warwick, UK.}
		\thanks{J.Gamper and S.Graham are also with the Mathematics for Real-World Systems Centre for Doctoral Training, University of Warwick, UK.}
		\thanks{M.Jahanifar is with the Department of Research and Development, NRP Co., Tehran, Iran}
		\thanks{K.Benes is with the Department of Pathology, Royal Wolverhampton NHS Trust, UK}
		\thanks{S.A.Khurram is with the Department of Clinical Dentistry, University of Sheffield, UK}
		\thanks{A.Azam and K.Hewitt are with the Department of Pathology at University Hospitals Coventry and Warwickshire, Coventry, UK}
		}

\maketitle
\thispagestyle{fancy}

\begin{abstract}
   The emerging area of computational pathology (CPath) is ripe ground for the application of deep learning (DL) methods to healthcare due to the sheer volume of raw pixel data in whole-slide images (WSIs) of cancerous tissue slides. However, it is imperative for the DL algorithms relying on nuclei-level details to be able to cope with data from `the clinical wild', which tends to be quite challenging.
   We study, and extend recently released PanNuke dataset consisting of nearly 200,000 nuclei categorized into 5 clinically important classes for the challenging tasks of segmenting and classifying nuclei in WSIs. Previous pan-cancer datasets consisted of only up to 9 different tissues and up to 21,000 unlabeled nuclei \cite{kumar2019multi} and just over 24,000 labeled nuclei with segmentation masks \cite{graham2019hover}. PanNuke consists of 19 different tissue types that have been semi-automatically annotated and quality controlled by clinical pathologists, leading to a dataset with statistics similar to `the clinical wild' and with minimal selection bias. We study the performance of segmentation and classification models when applied to the proposed dataset and demonstrate the application of models trained on PanNuke to whole-slide images. We provide comprehensive statistics about the dataset and outline recommendations and research directions to address the limitations of existing DL tools when applied to real-world CPath applications. 
\end{abstract}

\section{Introduction}

The success of convolutional neural networks (CNNs) in computer vision (CV) algorithms applied to natural image and medical imaging tasks can generally be attributed to the availability of large datasets and computing power \cite{campanella_terabyte-scale_2018, krizhevsky_imagenet_2012, liu_detecting_2017, sirinukunwattana_gland_2016, schaumberg_large-scale_2018}. Given the excellent performance on these tasks, measured by \textit{an average} metric evaluated over a large dataset, CNNs have sparked hope and promise in healthcare applications \cite{esteva_guide_2019, nagpal_development_2019}. 

The field of computational pathology (CPath) is witnessing a rapid rise in the research and development of deep learning (DL) models for quantitative profiling of spatial patterns in digitized whole-slide images (WSIs) of cancerous tissue slides that are rich in content and information \cite{colling2019artificial}. Numerous studies have demonstrated the potential of deep learning (DL) models in detecting cancer, classifying tissue, identifying diagnostically relevant structures and even inferring genetic sub-types \cite{shaban_prognostic_2018, javed_cellular_2018, nagpal_development_2019, fu_pan-cancer_2019, saltz_spatial_2018, sirinukunwattana_image-based_2019}. 

It would be fair to state that challenge contests have become a popular mean for attracting attention to a particular dataset or a task in medical imaging. In computational pathology, for instance, a variety of deep CNNs for automated nucleus segmentation have been developed on a dataset consisting of 21,623 nuclei in a total of 32 images of the size $1,000{\times}1,000$ pixels and released as part of the MoNuSeg challenge contest \cite{kumar2019multi}. However, the use and validity of results in most challenge contests is questionable due to the limited diversity\cite{reinke2018winner}. PanNuke, released under the CC license is however a diverse pan-cancer dataset that has undergone clinical quality control (QC).

\begin{figure}
\centering
\includegraphics[width=0.9\linewidth]{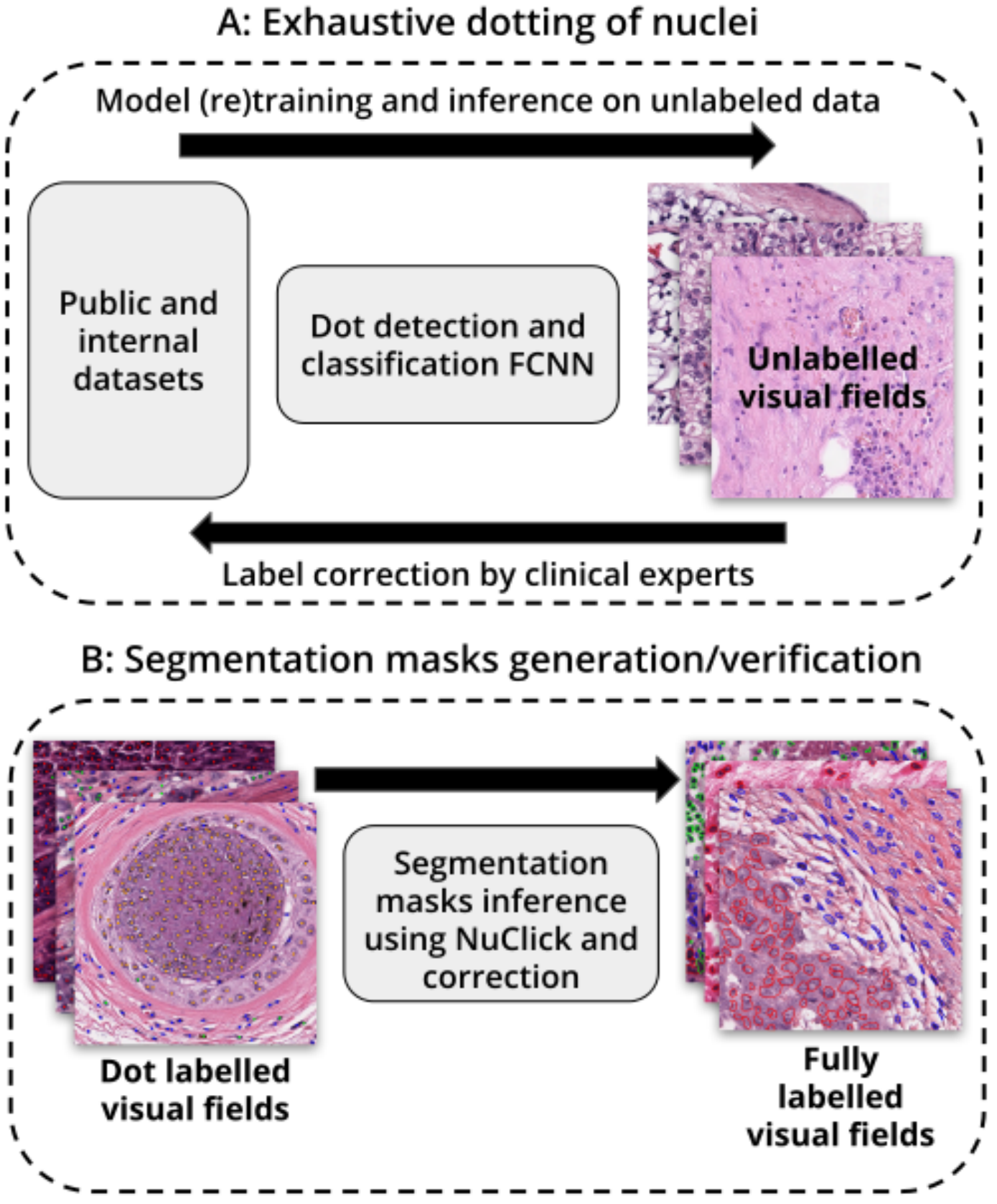}
   \caption{Illustration of PanNuke label generation and verification.}
\label{fig:illustration}
\end{figure}

\begin{figure*}
\begin{center}
  \includegraphics[width=1\linewidth]{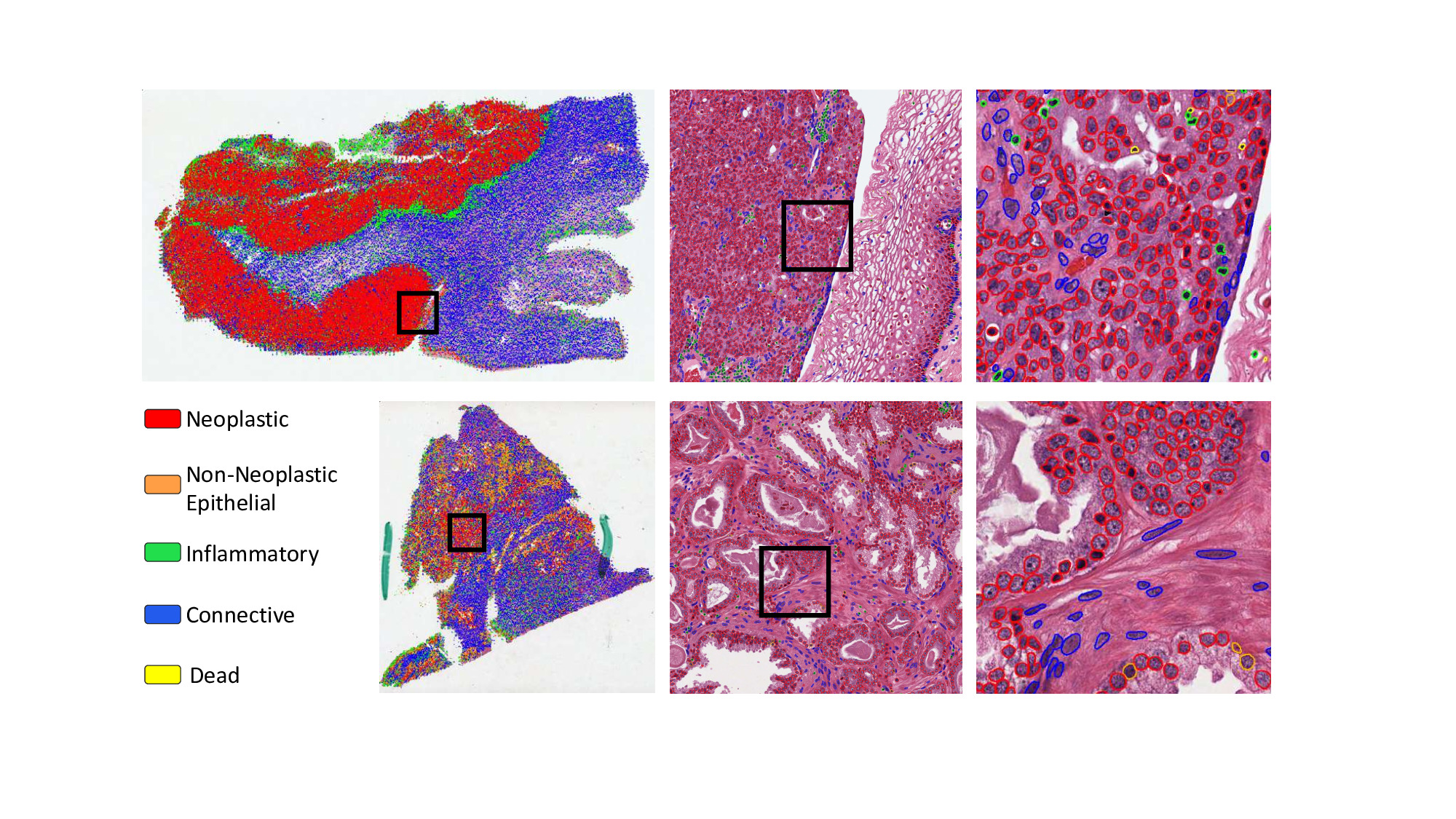}
\end{center}
  \caption{Visualisation of applying nuclei segmentation and classification network trained on PanNuke to unseen whole-slide images. Top row: Cervix tissue, with a visible differentiation between tumor and other tissue types. Bottom row: Prostate tissue.}
\label{fig:wsi_vis}
\end{figure*}

Second, CNN models applied to medical images are more powerful than the datasets that we apply them to, these models are known to suffer from over-fitting to surface statistical regularities \cite{brendel_approximating_2018, geirhos_generalisation_2018, jo_measuring_2017}. A model can obtain stronger inductive biases (a set of assumptions that a model uses to predict on a certain task) through multi-task learning, and therefore be more robust in practice \cite{baxter_model_2000}. The current approach in computational pathology is to train a model for a specific tissue, or even just a specific disease sub-type classification. On the other hand, we argue for a top-down algorithm development in medical imaging i.e. develop general and user-friendly tools that can easily be used by domain experts, and for bottom-up approach for dataset creation for supervised learning. Far too much medical imaging community's effort is focused on semi-supervised and unsupervised studies, however, these are far from clinically applicable and would always under-perform compared to supervised approaches trained on accurately labeled ground truth data. Models trained on PanNuke, as we demonstrate in Figure \ref{fig:wsi_vis}, could be used to assist detailed semi-automatic labeling in 19 different tissues. The results of nuclei detection and classification can be used for tissue pheno-typing as demonstrated in Colon tissue by Javed \textit{et al.} \cite{javed_cellular_2018}.

\begin{figure}
   \includegraphics[width=1\linewidth]{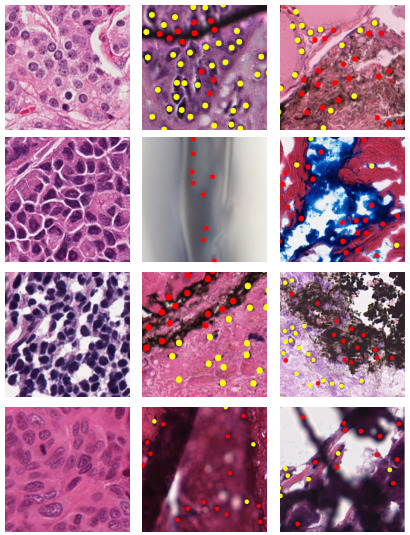}
   \caption{1st column: A selection of visual fields from the Kumar dataset \textit{et al.}\cite{kumar_dataset_2017}. 2nd-3rd columns: A selection of visual fields in PanNuke with output of a detector trained on \cite{kumar_dataset_2017} overlaid on the images. False positives (shown as red dots, as opposed to true positives as yellow dots) are clearly visible in the areas of burnt tissue, blur or other tissue processing or scanning artifacts.}
\label{fig:mistakes_misclass}
\end{figure}

This work is motivated by the fact that publicly available nucleus segmentation and classification datasets do not often match the distribution of data in `the clinical wild', as can be seen in Figure \ref{fig:mistakes_misclass} which shows the results of a nucleus detection model \cite{gamper_pannuke:_2019} trained on the MoNuSeg challenge dataset \cite{kumar2019multi}. It can be observed that when the model is applied to a few images that contain commonly found artifacts in clinical practice, there are several false detections which could lead to incorrect or misleading results in downstream CPath analysis. A similar phenomenon has been demonstrated by Oakden-Rayner \textit{et al.}\cite{oakden-rayner_exploring_2019} in the literature and in a validation study of DL  models  applied  to  radiology  images where model  performance  dropped  significantly  in  a  real-world  environment \cite{yune2018radiology}.

The main contributions of this work are summarized as below: 
\begin{itemize}
  \item We present PanNuke\footnote{Download dataset here: \url{https://warwick.ac.uk/fac/sci/dcs/research/tia/data/pannuke}}, the largest and the most diverse to date dataset for nucleus segmentation and classification, that has been annotated in a semi-automated manner and quality-controlled by clinical professionals;
  \item We accelerate the process of verification (i.e., quality control) by the clinical professionals by incorporating NuClick \cite{jahanifar_nuclick:_2019} during the generation of segmentation mask, as shown in Figure \ref{fig:illustration};
  \item We evaluate the performance of several nucleus segmentation models on PanNuke, which is 26 times larger in terms of unique 224$\times$224 training patches than the previous MoNuSeg challenge dataset \cite{kumar2019multi};
  \item We provide a full schema that could be used by the algorithm developers and that applies to other tissues not included in the PanNuke dataset. We show that segmentation models trained on PanNuke generalize to tissues like brain that were not part of the dataset.
  \item By releasing PanNuke to the broader CV research community, we encourage the community to develop new DL models to help push forward clinically relevant research.
\end{itemize}

These are not the only goals of this research. In fact, we hope to draw attention to the established tendencies in the field and their systemic impact on the risks as well as positive outcomes of CV in the healthcare domain and its progress.

In the following sections, we describe the relevant literature, our methodology, the quality of the automatically generated nucleus segmentation masks, provide qualitative and quantitative analysis of the dataset and it's significance. Finally, we evaluate the performance of existing approaches to nucleus segmentation and classification.

\subsection{Related Work}

Recent work in computational pathology has demonstrated that nuclear features can be effectively used for: cancer scoring, bio-marker discovery, cancer recurrence prediction  as well as for predicting treatment effectiveness \cite{beck_systematic_2011, chang_invariant_2013, sethi_abstract_2015, lee_nuclear_2017}. However, these are small sample size studies limited to a single tissue, and are focused on commonly studied tissues such as lung, breast, prostate and colon, due to the lack of data for other tissues. Javed \textit{et al.}\cite{javed_cellular_2018} demonstrated that inferred nuclear categories may help in classifying different tissue phenotypes within colon slides by constructing a graph of nuclei connections and detecting communities. By training models on PanNuke and applying them to WSIs at scale, these studies could be extended to 19 different tissues.

Closest in scale and real-world proximity to PanNuke is the work of Hosseini \textit{et al.} \cite{hosseini_atlas_nodate}, who have provided multi-category patch-level annotations. Due to having only patch annotations, Chan \textit{et al.}\cite{chan2019histosegnet} pursued semantic segmentation using class activation maps that significantly under-performed pixel-wise supervised models on the GlaS challenge dataset \cite{sirinukunwattana2017gland}. 
Across the board, large CPath studies have been somewhat limited by the lack of granular annotations \cite{fu2019pan, kather2019pan, campanella_terabyte-scale_2018}. 
However, PanNuke provides pixel-level boundary annotation for every individual nucleus, a building block of any organ's tissue. As a result, using models trained on PanNuke further semi-automatic labeling of tumor or tissue phenotypes is feasible \cite{javed_cellular_2018} (Figure \ref{fig:wsi_vis}).


\begin{table*}[h!]
\centering
\caption{Data used to initialize semi-automatic labeling.}
\begin{tabular}{lccccccc}
\hline \hline
\multirow{2}{*}{} & \multicolumn{5}{c}{\textbf{PanNuke Nuclei Categories}}                                                                                                   &       \\ \cline{2-8} 
                  & \textbf{Neoplastic} & \textbf{Non-Neo Epithelial} & \textbf{Inflammatory}  & \textbf{Connective} & \textbf{Dead}  & \multicolumn{1}{l|}{Non-Nuclei} & Total \\ \cline{2-8} 
MoNuSeg         & 5,927 & 836 & 1,698   & 906 & 0 & 0  & 9,367  \\
Colon Nuclei  & 4,685 & 7,544 & 6,003  & 4,468  & 2,547 & 0 & 25,247 \\
BreastPathQ        & 9,802 & 0 & 2,139  & 0 & 0 & 0 & 11,941 \\
Nuclei Attribute   & 0  & 0  & 0 & 0 & 0 & 500 & 500   \\ \hline
Total             & 20,414 & 8,380 & 9,840 & 5,374 & 2,547       & 500 & 47,055 \\ \hline \hline
\end{tabular}
\label{table:init_data}
\end{table*}

\section{The PanNuke Dataset}

In the sections below, we describe our methodology, discuss the quality of the automatically generated nucleus segmentation masks and provide qualitative and quantitative analysis of the dataset and its significance in algorithmic and practical terms. 

\begin{figure*}[ht!]
\centering
\includegraphics[width=0.8\linewidth]{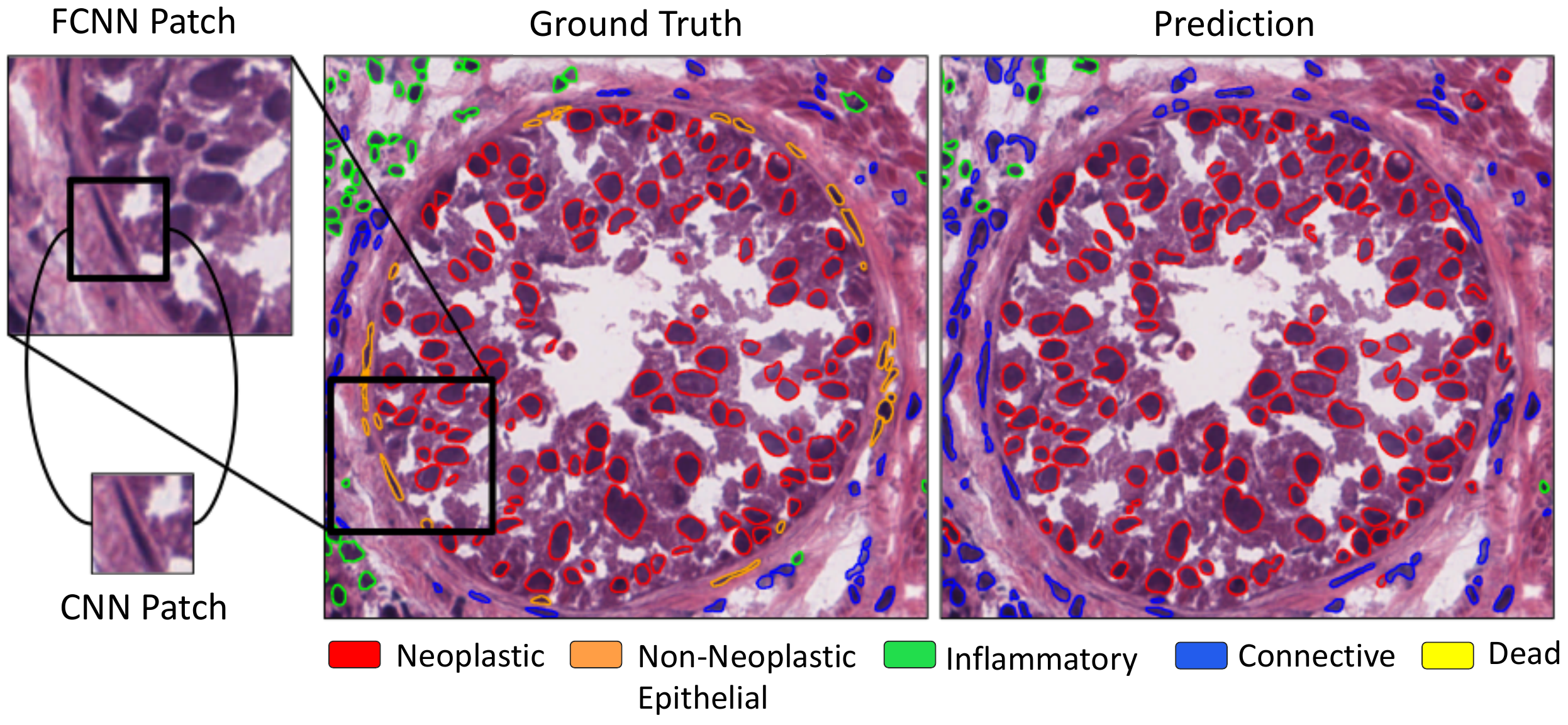}
   \caption{Ground truth labels and segmentation masks verified by pathologists alongside model prediction for bladder tissue visual field. FCNN patch represents a 224${\times}$224 patch commonly used for training fully convolution segmentation models, right below it is a patch used by \cite{sirinukunwattana_locality_2016} for training a CNN to classify each individual nucleus.}
\label{fig:bladder_images}
\end{figure*}

\subsection{Dataset Generation}
\begin{figure}{}
\centering
\includegraphics[width=0.95\linewidth]{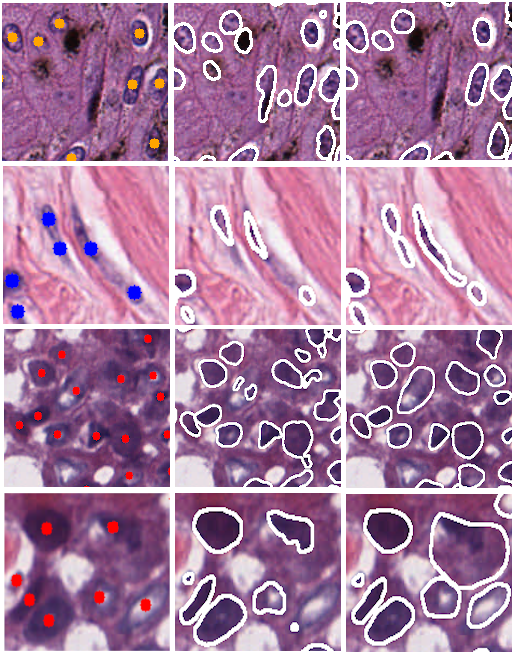}
   \caption{Left column: Pathologist verified nuclei dots; Middle: CNN generated segmentation masks; Right: NuClick generated segmentation masks and subsequently verified. Color dots are consistent with the legend in Figure \ref{fig:bladder_images}.}
\label{fig:nuclick_vs_cnn}
\end{figure}
\textbf{Data Labeling:} First, we aggregated a set of publicly available nucleus classification and detection datasets to create an initial dataset for semi-automatic ground truth generation. For this, we trained a fully convolutional neural network (FCNN) for nucleus detection using 4 publicly available datasets: Kumar~\cite{kumar_dataset_2017}, CPM2017~\cite{vu2019methods}, 15 visual field from TCGA \cite{liu2018integrated} that we have labeled ourselves, and a dataset of bone marrow visual fields \cite{kainz_you_2015}. Kumar is a dataset of 16 visual fields consisting of 7 different tissue types (liver, prostate, kidney, breast, stomach, colorectal and bladder) and CPM2017 is a dataset of 64 visual fields of 4 cancer types (glioblastoma, low grade glioma, head neck squamous cell carcinoma and non-small cell lung cancer). Therefore, in total we initially utilize 106 visual fields. We then trained a convolutional neural network on nuclei CNN patches (see Figure \ref{fig:bladder_images} for illustration) extracted from the datasets described in Table \ref{table:init_data}\footnote{BrestPathQ source: \url{https://breastpathq.grand-challenge.org}}, that were re-labeled according to the PanNuke categories. The schema for nuclei categories will be described in the subsequent sections. Using the CNN, we exhaustively classified all of the detected nuclei in the above mentioned datasets, which were then verified by a team of expert pathologists. After, we sampled 2,000 visual fields from more than 20,000 WSIs in 19 different tissues obtained from TCGA and also from a local hospital. Random sampling of visual fields allowed us to address the selection bias present in available public datasets, in fact visual fields with common clinical artifacts demonstrated in Figure \ref{fig:mistakes_misclass} are from the final PanNuke dataset.

When sampling visual fields of tissue from WSIs, the tissue may have been originally frozen or paraffin embedded and also the WSIs may have been scanned with a maximum resolution of either 20$\times$ or 40$\times$. We re-sized the selected visual fields so that all images were at 40$\times$ resolution and we excluded frozen tissue from the study. We then re-sampled the visual fields so that the images present in the dataset were diagnostically relevant and also reflected the true variation of tissue in each organ. We also ensured that artifacts remained in the dataset because they are inevitable when facing data in the `clinical wild'.  

We then proceeded to Part A of annotation process, as illustrated in Figure \ref{fig:illustration}. Iterating 7 times, where after each stage pathologists would verify and re-label the detected and classified nuclei dots. In each stage, the FCNN detection and classification model was trained with the new collected annotations to provide better predictions for next iteration. Eventually, this leads to a dataset of 481 visual fields with a total of 189,744 exhaustively annotated nuclei, verified by domain experts.

\begin{figure*}
   \includegraphics[width=1\linewidth]{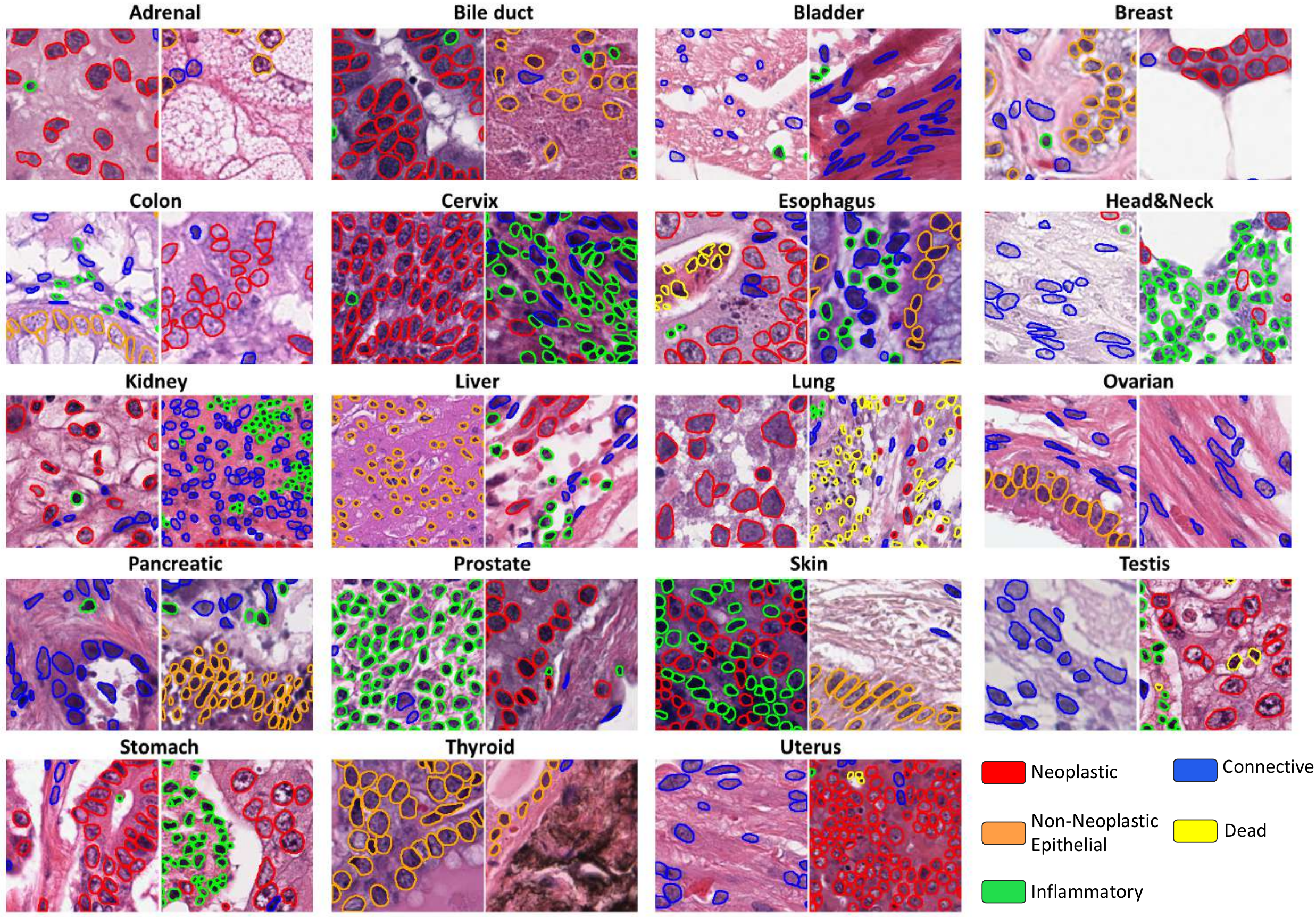}
   \caption{Example of PanNuke patches and their ground truth annotations overlaid.}
\label{fig:patches_vis}
\end{figure*}
\textbf{Mask Generation:} For the final version of PanNuke we used NuClick \cite{jahanifar_nuclick:_2019}. Nuclick is a method that enables accurate segmentation mask generation from a single point. Therefore, this enables us to produce many segmentation masks at a relatively low cost and also eases the verification process because only a single point is required from the pathologist, as opposed to the entire mask. Figure \ref{fig:nuclick_vs_cnn} presents a selection of masks generated by a segmentation FCNN, as well as by NuClick. In the first row, the FCNN has mistaken pigment in skin tissue as nuclei. However, because NuClick is conditioned on verified nuclei dots, only nuclei pixels are segmented. Also, the FCNN frequently misses elongated nuclei compared to NuClick and often struggles to segment nuclei with indistinct boundaries, as pictured in the second column and third row. Finally, the last row shows that FCNN only segments the nucleolus, whereas NuClick segments the entire nucleus.

Using the proposed semi-automatic pipeline, we generated and verified 189,744 nuclei from more than 20,000 WSIs. Some examples of generated ground truth are depicted in the Figure \ref{fig:patches_vis}.
\begin{figure}{}
\centering
   \includegraphics[width=1\linewidth]{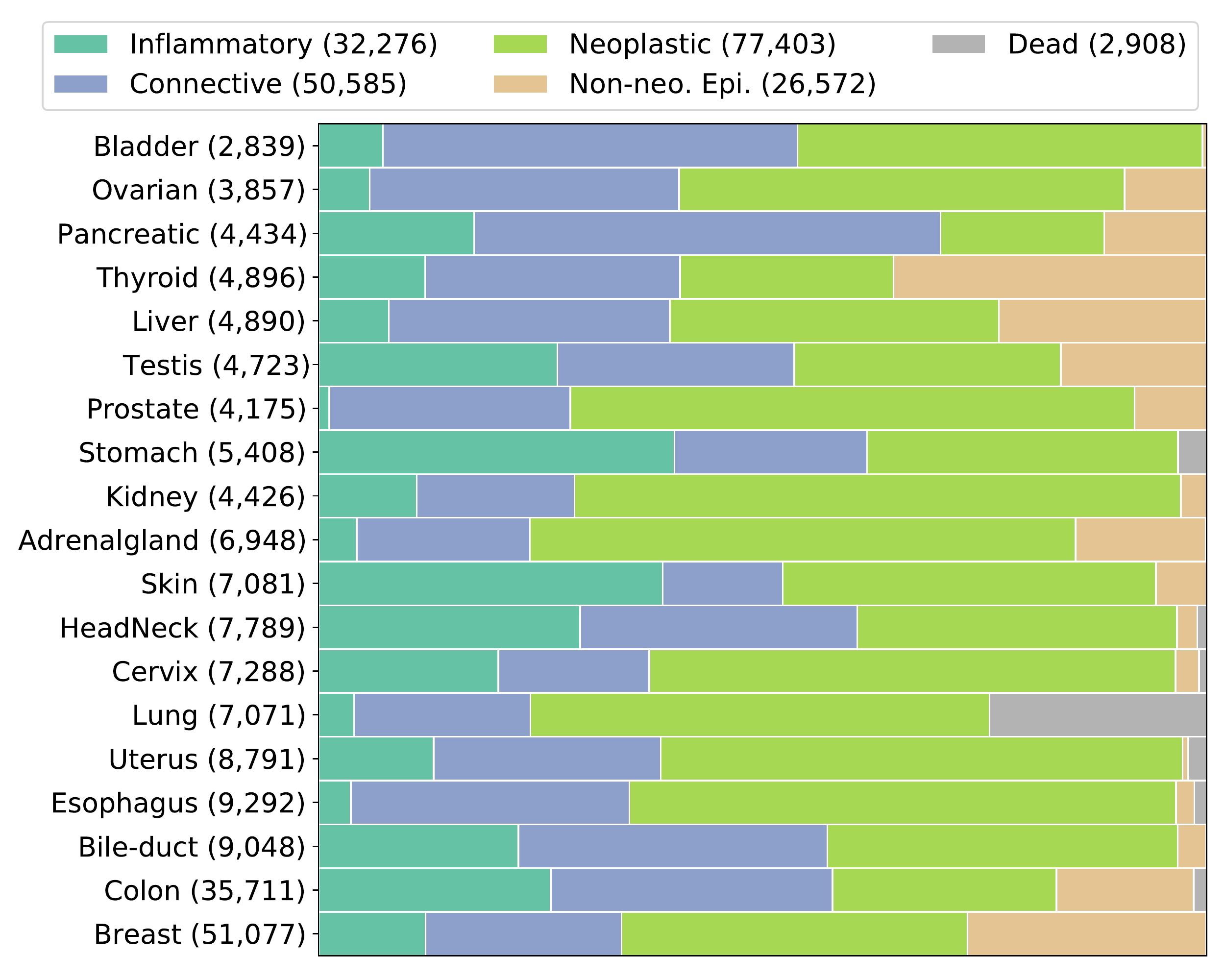}
   \caption{A comparative plot of class distributions per tissue. Numbers in parenthesis represent the total number of nuclei within that category or tissue type.}
\label{fig:nuclei_count}
\end{figure}
\subsection{Dataset Description and Statistics}
\label{analysis}
\textbf{Dataset Schema:} For the purpose of this work, we derived a \textit{schema} in Table \ref{table:full_schema} for generating nuclei labels that is clinically sound and is shared across the 19 tissues within the dataset. The proposed schema provides insight into how we categorized all nuclei and can be used to appropriately sub-type nuclei in future studies. Our schema is consistent with the nuclei categories used in previous tissue-specific studies \cite{sirinukunwattana_locality_2016, graham2019hover, hosseini_atlas_nodate} and therefore work previously developed on these datasets can be seamlessly extended to PanNuke. We split the cell types between Neoplastic and Non-neoplastic cells. 

\begin{table}[t!]
\centering
\caption{Nucleus classification schema for computational pathology.}
\label{table:full_schema}
\begin{tabular}{l!{\vrule width \lightrulewidth}l!{\vrule width \lightrulewidth}l!{\vrule width \lightrulewidth}l} 
\hline\hline
\multicolumn{1}{l}{\#} & \multicolumn{1}{l}{\textbf{Level-1 } }                                                                             & \multicolumn{1}{l}{\textbf{Level-2 } }                                                  & \textbf{Level-3 }                                                                      \\ 
\hline
1                      & \multirow{3}{*}{Epithelial}                                                                                        & \multirow{2}{*}{Neoplastic}                                                             & Malignant\textasciitilde{}                                                             \\ 
\cline{1-1}\cline{4-4}
2                      &                                                                                                                    &                                                                                         & Benign\textasciitilde{}                                                                \\ 
\cline{1-1}\cline{3-4}
3                      &                                                                                                                    & \multicolumn{2}{l}{Non neoplastic}                                                                                                                                               \\ 
\hline
4                      & \multirow{5}{*}{\begin{tabular}[c]{@{}l@{}} Connective/\\Soft tissue cells \end{tabular}} & \multicolumn{2}{r}{Fibroblasts}                                                                                                                                                  \\ 
\cline{1-1}\cline{3-4}
5                      &                                                                                                                    & \multicolumn{2}{r}{Endothelial}                                                                                                                                                  \\ 
\cline{1-1}\cline{3-4}
6                      &                                                                                                                    & \multicolumn{2}{r}{Myo-fibroblasts}                                                                                                                                              \\ 
\cline{1-1}\cline{3-4}
7                      &                                                                                                                    & \multicolumn{2}{r}{Fibers}                                                                                                                                                       \\ 
\cline{1-1}\cline{3-4}
8                      &                                                                                                                    & \multicolumn{2}{r}{Adipocytes}                                                                                                                                                   \\ 
\hline
9                      & \multirow{9}{*}{\begin{tabular}[c]{@{}l@{}} Lympho-reticular \\cells \end{tabular}}       & \multicolumn{2}{r}{Trombocytes}                                                                                                                                                  \\ 
\cline{1-1}\cline{3-4}
10                     &                                                                                                                    & \multicolumn{2}{r}{Erythrocytes}                                                                                                                                                 \\ 
\cline{1-1}\cline{3-4}
11                     &                                                                                                                    & \multirow{3}{*}{Leukocytes}                                                             & Eosinophils                                                                            \\ 
\cline{1-1}\cline{4-4}
12                     &                                                                                                                    &                                                                                         & Basophils                                                                              \\ 
\cline{1-1}\cline{4-4}
13                     &                                                                                                                    &                                                                                         & Neutrophils                                                                            \\ 
\cline{1-1}\cline{3-4}
14                     &                                                                                                                    & \multirow{2}{*}{Lymphoid cells}                                                         & Plasma cells                                                                           \\ 
\cline{1-1}\cline{4-4}
15                     &                                                                                                                    &                                                                                         & Lymphocytes                                                                            \\ 
\cline{1-1}\cline{3-4}
16                     &                                                                                                                    & \multicolumn{2}{r}{Macrophages/Histiocytes}                                                                                                                                       \\ 
\cline{1-1}\cline{3-4}
17                     &                                                                                                                    & \multicolumn{2}{r}{Mast cells}                                                                                                                                                   \\ 
\hline
18                     & \multirow{6}{*}{\begin{tabular}[c]{@{}l@{}} Nervous system \\cells \end{tabular}}         & \multirow{4}{*}{CNS}                                                                    & \begin{tabular}[c]{@{}l@{}}Oligo-\\dendrocytes \end{tabular}  \\ 
\cline{1-1}\cline{4-4}
19                     &                                                                                                                    &                                                                                         & Microglia                                                                              \\ 
\cline{1-1}\cline{4-4}
20                     &                                                                                                                    &                                                                                         & Astrocyte                                                                              \\ 
\cline{1-1}\cline{4-4}
21                     &                                                                                                                    &                                                                                         & \begin{tabular}[c]{@{}l@{}}Ependymal \\cells \end{tabular}    \\ 
\cline{1-1}\cline{3-4}
22                     &                                                                                                                    & \begin{tabular}[c]{@{}l@{}}Peripherial \\Nervous \end{tabular} & \begin{tabular}[c]{@{}l@{}}Schwann \\cells \end{tabular}      \\ 
\cline{1-1}\cline{3-4}
23                     &                                                                                                                    & Ganglia                                                                                 & \begin{tabular}[c]{@{}l@{}}Ganglion \\cells \end{tabular}     \\ 
\hline
24                     & \multirow{2}{*}{Dead}                                                                                              & \multicolumn{2}{r}{Apoptotic}                                                                                                                                                    \\ 
\cline{1-1}\cline{3-4}
25                     &                                                                                                                    & \multicolumn{2}{r}{Necrotic}                                                                                                                                                     \\
\hline
\end{tabular}
\end{table}

Neoplasm ('new growth') includes any tumor, malignant or benign. It includes carcinomas, sarcomas, melanomas, lymphomas, etc. These are all tumors but originate from different cell types: carcinomas from epithelial; sarcomas from soft tissue; melanomas from melanocytes; lymphomas from lymphoid cells and so on. As such, all tumorous cells in PanNuke are labeled as Neoplastic. 

Non-neoplastic covers everything else, from normal to inflammatory, degenerative, metaplastic, atypia etc. For the purpose of this exercise, atypia is under the heading of non-neoplastic, although some researchers may argue that they have clonal changes and potential for turning into neoplastic cells. As such, non-neoplastic labels in PanNuke are: epithelial; connective/soft tissue cells; inflammatory and dead cells. Here, connective tissue cells have the potential to become neoplastic, whereas inflammatory cells typically cannot become neoplastic. Inflammatory cells include lymphoid and macrophage cells in PanNuke. Dead cells can arise from either neoplastic or non-neoplastic cells, but in this study we refer to them as non-neoplastic.


\textbf{Dataset Statistics:} In general, most nuclei types that we provide in our schema are represented in all tissues considered in PanNuke, but the distribution of the nuclei count per class may vary from tissue to tissue. This can be seen in Figure \ref{fig:nuclei_count}, where we observe that the total nuclei count per tissue as well as per class  changes between tissue types.

In our experience, a pathologist when annotating nuclei frequently refers to the WSI at a lower resolution, to observe the surrounding structures. Seminal work in automatic nuclei classification by Sirinukunwattana \textit{et al.}\textit{et al.}\cite{sirinukunwattana_locality_2016} considered only nuclear patches when performing classification (CNN patch in Figure \ref{fig:bladder_images}), or small image patches containing a few nuclei in the more recent work using fully convolutional networks (FCNN patch in Figure \ref{fig:bladder_images}) - both approaches have demonstrated high accuracy on average. What does it mean for us? Either that pathologist overfits given the surrounding; or that looking at a nuclear patch to classify nuclei is not sufficient; or an \textit{average} accuracy metric is not a clinically relevant measure of algorithmic performance.  Recently Oakden \textit{et al.}\cite{oakden-rayner_hidden_2019} has demonstrated effects of \textit{hidden stratification} in medical imaging, i.e. when labels used in CV or ML studies do not represent clinical reality, and that \textit{average} performance is not a strong measure of applicability. As Figure \ref{fig:bladder_images} demonstrates \textit{hidden stratification} is also present in histology, if one is to use nuclei classification for downstream tasks. Carcinoma (malignant epithelial tumor/neoplasm) can be non invasive (i.e. in situ) or invasive. They are both composed of malignant cells but in situ is still "bounded" by either basal cell layer/myoepithelial layer or basement membrane (depending on organ site) hence called non-invasive.  Invasive carcinomas have lost basal cells/myoepithelial cells. So in theory, non invasive cancer should have no or low potential for lympho-vascular space invasion, distant metastases and similar (i.e. they should have better prognosis than or may not need as radical treatment as invasive carcinomas). For this specific image, for example, it would appear that this tumor cells are 'bounded' by basement membrane cells (Figure \ref{fig:bladder_images} Ground Truth). While FCNN based model (Figure \ref{fig:bladder_images} Prediction) classifies epithelial cells surrounding tumor cells as part of connective tissue category, evidently due to their shapes and the tendency of deep models to look for simplest association that describe the relationship between the feature and target on average \cite{brendel_approximating_2018, jo_measuring_2017}. Notably, these cells would be unlikely correctly classified by ANY algorithm. We either need to incorporate prior knowledge about the underlying mechanisms in the tissue in order to solve these tasks as precisely as a pathologist, or acquire multiple, similar cases and train on substantially larger image sizes than FCNN patch. 

These are just a few examples of cases which are difficult to simply classify/diagnose as neoplastic or non-neoplastic. Other challenging examples could not be included given the limited space, but in clinical practice these cases are more frequent as compared to how they are portrayed in the AI/digital pathology literature.

\section{Performance Benchmarks}

Previous nucleus segmentation datasets provided visual fields and therefore patch extraction between different methods for training and testing is not standardized. For PanNuke we pre-extract patches and split into 3 randomized training, validation and testing folds for a fair model comparison. For every fold we split every tissue into three sections by ensuring that each contains an equal portion of the smallest class within it (refer to Figure \ref{fig:nuclei_count}). Here, we apply recent and well known models on PanNuke to create a benchmark for further research using this dataset.

\begin{figure*}[h!]
\centering
  \includegraphics[width=0.9\linewidth]{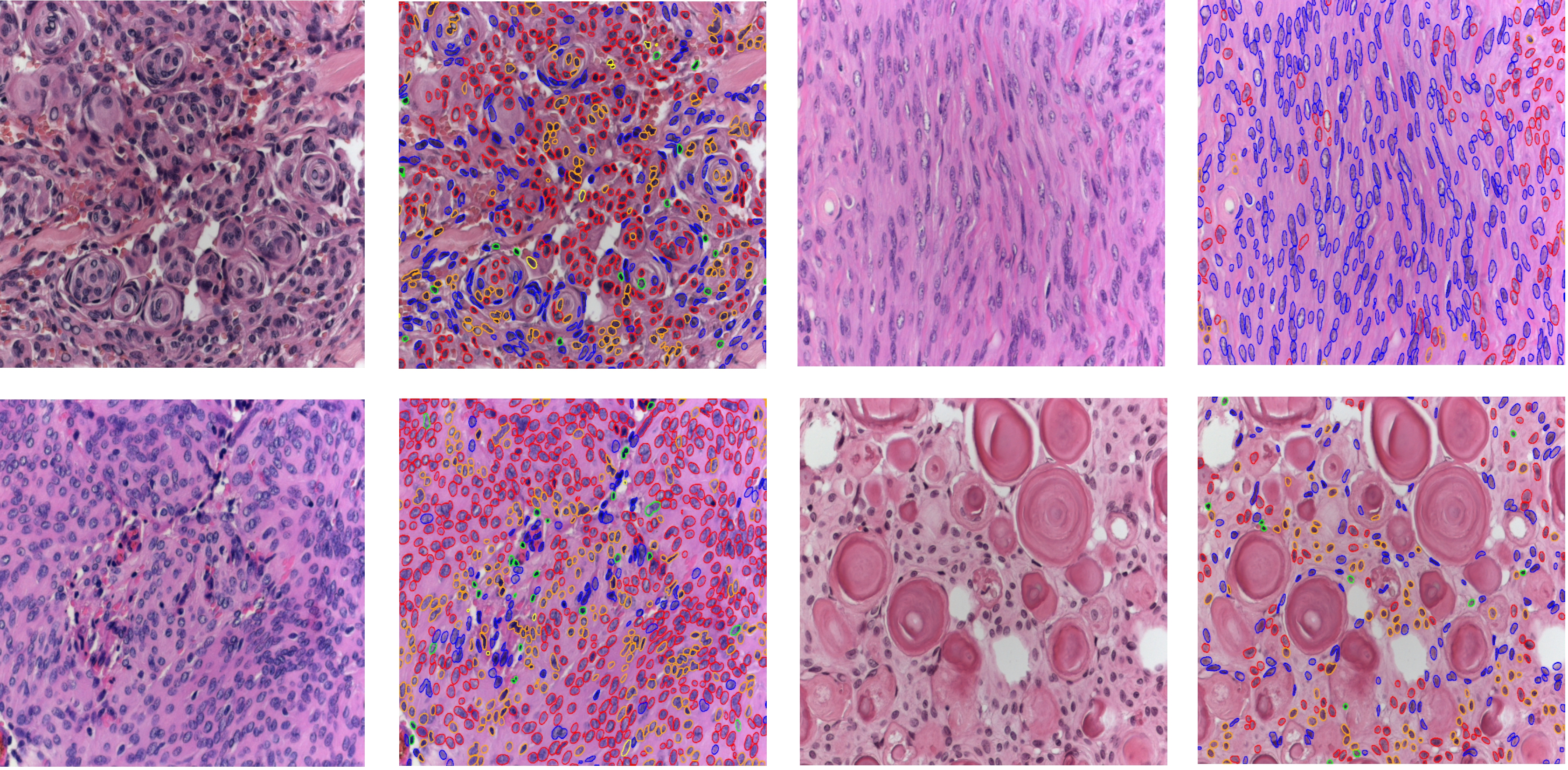}
  \caption{Brain tissue visual fields from Qureshi \textit{et al.}\cite{qureshi_adaptive_2008} and predictions overlay using HoVer-Net trained on PanNuke.}
\label{fig:brain}
\end{figure*}

\subsection{Baseline Models}

There are not an abundance of models that perform simultaneous segmentation and classification and therefore we adapt several top-performing instance segmentation models so that they additionally classify each nucleus.
We quantified the performance of 4 models on PanNuke: DIST \cite{naylor2018segmentation} which utilizes the distance map of instances as the target and we add another branch for semantic pixel-wise classification; Mask-RCNN \cite{he2017mask} which is a state-of-the-art instance segmentation network for natural images; Micro-Net \cite{raza2019micro} which 
was proposed for nuclear and gland segmentation and HoVer-Net \cite{graham2019hover}, which uses the concept of horizontal and vertical distance maps to separate clustered nuclei. HoVer-Net does not need to be adapted as it inherently performs simultaneous nuclear instance segmentation and classification. Note, that each of the above mentioned models performed extensive comparison with competing segmentation methods and therefore we focus on the best performing models.

In addition to the segmentation models, we also implemented a detection-based U-Net, similar to the model proposed by Xie \textit{et al.}~\cite{xie2018efficient}, where the detection maps for each class were used as the target for training the network. Specifically, the detection map for each nucleus was converted to a 2D Gaussian centered at the true nucleus centroid. 

\subsection{Evaluation}
\textbf{Instance Segmentation:} To quantify the instance segmentation performance of each of the models trained on PanNuke, we use panoptic quality (PQ) \cite{kirillov2019panoptic,graham2019hover}. An in depth discussion as to why we choose to utilize PQ over other recently used metrics for nucleus segmentation is discussed by Graham \textit{et al.} \cite{graham2019hover}. Specifically, we used multi-class PQ (mPQ) and binary PQ (bPQ) that assumes that all nuclei belong to one class. For mPQ, the PQ is calculated independently for each positive class and then the results are averaged. Therefore, the metric is insensitive to class imbalance. Our main criterion for evaluating model performance is the average mPQ over all of the tissues, which therefore equally weights the contribution of each tissue type. The IoU threshold for determining a true positive during PQ calculation is set to 0.5. As part of this work, we provide the implementation of our evaluation framework\footnote{Evaluation code: \url{https://github.com/TIA-Lab/PanNuke-metrics}} to encourage reproducibility, however this is a preliminary release of the code and we are working on a full repository release.

To provide insight into how each model performs for different types of nuclei, in Table \ref{table:TissuePQs}, we report mPQ and bPQ for all 19 tissue types separately. We observe that HoVer-Net achieves the best performance for most tissue types and is reflected by the greatest average score over all tissues for mPQ and bPQ. We also report PQ for each type of nucleus in Table \ref{table:PQtype}. Dead cells obtain a low PQ for all models because these nuclei are very small and therefore achieving an IoU$>$0.5 (PQ criterion for true positive) is difficult. In some cases, distinguishing between neoplastic and non-neoplastic nuclei proved to be challenging, yet it must be emphasized that this also can be a challenging task for the pathologist, where they often need to assess contextual information before confirming the the type of nucleus. As deduced from Figure \ref{fig:nuclei_count} class imbalance may also lead to poor performance for dead cell and non-neoplastic classes.

\textbf{Detection:} In order to allow cross comparison with detection models (as opposed to segmentation), we reported the F\textsubscript{1}, precision and recall for the overall detection quality in Table \ref{table:detection}. Here, a true positive was considered as a detection within 12 pixels of the labeled centroid \cite{sirinukunwattana_locality_2016}. We also calculated. In order to report the detection performance for segmentation models, we extracted the centroids of each instance as detection points. We observed that segmentation models generally performed better than the detection model. We hypothesize that this is because the detection-based model does not incorporate boundary information.

\begin{table*}[h!]
\centering
\caption{Average mPQ and bPQ across three dataset splits. We also provide the standard deviation (STD) across these splits in the final row.}
\label{table:TissuePQs}
\begin{tabular}{l|cc|cc|cc|cc}
\multicolumn{1}{c|}{}  & \multicolumn{2}{c|}{DIST} & \multicolumn{2}{c|}{Mask-RCNN} & \multicolumn{2}{c|}{Micro-Net} & \multicolumn{2}{c}{HoVer-Net}                            \\ 
\midrule
\multicolumn{1}{c|}{}   & mPQ & bPQ & mPQ & bPQ & mPQ  & bPQ  & mPQ  & bPQ  \\
Adrenal Gland  & 0.3442 & 0.5603 & 0.3470 & 0.5546 & 0.4153 & 0.6440 & 0.4812 & 0.6962 \\ 
\hdashline
Bile Duct & 0.3614  & 0.5384 & 0.3536 & 0.5567 & 0.4124  & 0.6232  & 0.4714  & 0.6696 \\ 
\hdashline
Bladder  & 0.4463 & 0.5625 & 0.5065 & 0.6049 & 0.5357 & 0.6488 & 0.5792 & 0.7031 \\ 
\hdashline
Breast & 0.3790 & 0.5466 & 0.3882 & 0.5574 & 0.4407 & 0.6029 & 0.4902 & 0.6470 \\ 
\hdashline
Cervix & 0.3371 & 0.5309 & 0.3402 & 0.5483 & 0.3795 & 0.6101 & 0.4438 & 0.6652 \\ 
\hdashline
Colon & 0.2989 & 0.4508 & 0.3122 & 0.4603 & 0.3414 & 0.4972 & 0.4095 & 0.5575  \\ 
\hdashline
Esophagus  & 0.3942 & 0.5295 & 0.4311  & 0.5691 & 0.4668 & 0.6011 & 0.5085 & 0.6427 \\ 
\hdashline
Head \& Neck & 0.3177 & 0.4764 & 0.3946 & 0.5457 & 0.3668 & 0.5242 & 0.4530 & 0.6331 \\ 
\hdashline
Kidney & 0.3339 & 0.5727 & 0.3553 & 0.5092 & 0.4165 & 0.6321 & 0.4424 & 0.6836 \\ 
\hdashline
Liver  & 0.3441 & 0.5818 & 0.4103 & 0.6085 & 0.4365 & 0.6666 & 0.4974 & 0.7248 \\
\hdashline
Lung  & 0.2809 & 0.4978 & 0.3182 & 0.5134 & 0.3370 & 0.5588 & 0.4004  & 0.6302 \\ 
\hdashline
Ovarian & 0.3789 & 0.5289 & 0.4337 & 0.5784 & 0.4387 & 0.6013 & 0.4863 & 0.6309 \\ 
\hdashline
Pancreatic  & 0.3395 & 0.5343 & 0.3624 & 0.5460  & 0.4041 & 0.6074 & 0.4600 & 0.6491 \\ 
\hdashline
Prostate & 0.3810 & 0.5442 & 0.3959 & 0.5789 & 0.4341 & 0.6049 & 0.5101 & 0.6615 \\ 
\hdashline
Skin    & 0.2627 & 0.5080 & 0.2665 & 0.5021 & 0.3223 & 0.5817 & 0.3429 & 0.6234 \\ 
\hdashline
Stomach & 0.3369 & 0.5553  & 0.3684 & 0.5976 & 0.3872  & 0.6293 & 0.4726  & 0.6886 \\ 
\hdashline
Testis & 0.3278 & 0.5548 & 0.3512 & 0.5420 & 0.4088 & 0.6300 & 0.4754 & 0.6890 \\ 
\hdashline
Thyroid  & 0.2574 & 0.5596 & 0.3037 & 0.5712 & 0.3712 & 0.6555  & 0.4315 & 0.6983 \\ 
\hdashline
Uterus & 0.3487 & 0.5246 & 0.3683 & 0.5589 & 0.3965 & 0.5821  & 0.4393 & 0.6393 \\ 
\hline
Average across tissues & \textbf{0.3406} & \textbf{0.5346} & \textbf{0.3688} & \textbf{0.5528} & \textbf{0.4059} & \textbf{0.6053} & \textbf{0.4629} & \textbf{0.6596}  \\
STD across splits  & 0.0156 & 0.00975 & 0.00465 & 0.00762 & 0.00816 & 0.00499 & 0.00758 & 0.00364
\end{tabular}
\end{table*}


\begin{table}
\centering
\caption{Average PQ across three dataset splits for each nuclear category.}
\begin{tabular}{lccccc}
          & \multicolumn{1}{l}{Neo} & \multicolumn{1}{l}{Non-Neo Epi~} & \multicolumn{1}{l}{Inflam~} & \multicolumn{1}{l}{Conn~} & \multicolumn{1}{l}{Dead~~}  \\ 
\hline
DIST      & 0.439 & 0.290 & 0.343   & 0.275 & 0.000                       \\ 
\hdashline
Mask-RCNN & 0.472  & 0.403 & 0.290 & 0.300 & 0.069                        \\ 
\hdashline
Micro-Net & 0.504 & 0.442 & 0.333 & 0.334 & 0.051                       \\ 
\hdashline
HoVer-Net & \textbf{0.551} & \textbf{0.491} & \textbf{0.417} & \textbf{0.388} & \textbf{0.139}                       
\end{tabular}
\label{table:PQtype}
\end{table}

\subsection{Generalisation to other tissues}

We speculated that models trained on PanNuke would likely generalise to other tissues and applied the best performing model to brain tissue as demonstrated in Figure \ref{fig:brain}. Within this tissue, the model was able to perform a successful segmentation of all nuclei, but found it challenging to predict the correct nuclear categories. The algorithm trained with PanNuke performs favourably for segmentation of nuclei for the 4 images in Figure S3  (DICE value 0.796, mPQ 0.28 and bPQ 0.51) from a completely unseen source (Germany) and tissue type (brain), as compared to the tissue-wise average in Table 3.


\section{Concluding Remarks}
In this paper, we presented a semi-annotated and quality-controlled dataset with detailed boundaries and class labels for 5 main types of nuclei for multiple different cancerous tissue types.
This work is motivated by the observation that the use and validity of results in most challenge contests is questionable due to the limited nature of challenge datasets \cite{reinke2018winner}. For example, even on ImageNet \cite{imagenet_cvpr09} (which happens to be many orders or magnitude larger than \cite{kumar_dataset_2017}), there is evidence of overfitting due to multiple-hypothesis testing in architecture development \cite{werpachowski_detecting_2019}.
In addition to selection bias in the available datasets, results are reported on labels that do not always meaningfully describe the variation within the population.
This work, while providing a significant contribution in modeling and dataset size compared to any previous work, is only a small step in the direction of safe and robust application of CV in CPath. Similarly to Esteva \textit{et al.}\cite{esteva_dermatologist-level_2017}, we offer a careful treatment of PanNuke labels, discuss the real world complexities of the task and offer \textit{schemas} that researchers can use to push nuclei classification research further.

\begin{table*}[h]
\centering
\caption{Precision (P), Recall (R) and F\textsubscript{1} score for detection and classification. All results are the average across three dataset splits. For segmentation predictions, the centroid of each nucleus is extracted for computing these metrics.}
\begin{tabular}{lccc!{\vrule width \lightrulewidth}ccc!{\vrule width \lightrulewidth}ccc!{\vrule width \lightrulewidth}ccc!{\vrule width \lightrulewidth}ccc!{\vrule width \lightrulewidth}ccc} 
\toprule
\multicolumn{4}{c!{\vrule width \lightrulewidth}}{\multirow{2}{*}{Detection}}                                                        & \multicolumn{15}{c}{Classification}                                                                                                            \\ 
\cmidrule[\heavyrulewidth]{5-19}
\multicolumn{4}{c!{\vrule width \lightrulewidth}}{}                                                                                  & \multicolumn{3}{c!{\vrule width \lightrulewidth}}{Neoplastic}                                                 & \multicolumn{3}{c!{\vrule width \lightrulewidth}}{Non-Neo Epithelial}                                             & \multicolumn{3}{c!{\vrule width \lightrulewidth}}{Inflammatory}                                               & \multicolumn{3}{c!{\vrule width \lightrulewidth}}{Connective}                                                 & \multicolumn{3}{c}{Dead}                                                  \\ 
\toprule
\multicolumn{1}{c}{} & P  & R  & Fd  & P  & R  & F1   & P & R & F1 & P & R & F1   & P    & R  & F1 & P & R  & F1   \\ 
\hdashline
Det U-Net            & \multicolumn{1}{l}{0.73} & \multicolumn{1}{l}{0.59} & \multicolumn{1}{l!{\vrule width \lightrulewidth}}{0.65} & \multicolumn{1}{l}{0.40} & \multicolumn{1}{l}{0.47} & \multicolumn{1}{l!{\vrule width \lightrulewidth}}{0.43} & \multicolumn{1}{l}{0.27} & \multicolumn{1}{l}{0.31} & \multicolumn{1}{l!{\vrule width \lightrulewidth}}{0.29} & \multicolumn{1}{l}{0.32} & \multicolumn{1}{l}{0.45} & \multicolumn{1}{l!{\vrule width \lightrulewidth}}{0.37} & \multicolumn{1}{l}{0.34} & \multicolumn{1}{l}{0.38} & \multicolumn{1}{l!{\vrule width \lightrulewidth}}{0.36} & \multicolumn{1}{l}{0.00} & \multicolumn{1}{l}{0.00} & \multicolumn{1}{l}{0.00}  \\ 
\hdashline
DIST                 & 0.74                     & 0.71                     & 0.73                                                    & 0.49                     & 0.55                     & 0.50                                                    & 0.38                     & 0.33                     & 0.35                                                    & 0.42                     & 0.45                     & 0.42                                                    & 0.42                     & 0.37                     & 0.39                                                    & 0.00                     & 0.00                     & 0.00                      \\ 
\hdashline
MRCNN                & 0.76                     & 0.68                     & 0.72                                                    & 0.55                     & 0.63                     & 0.59                                                    & 0.52                     & 0.52                     & 0.52                                                    & 0.46                     & 0.54                     & 0.50                                           & 0.42                     & 0.43                     & 0.42                                                    & 0.17                     & 0.30                     & 0.22                      \\ 
\hdashline
Micro-net            & 0.78                     & 0.82                     & \textbf{0.80}                                           & 0.59                     & 0.66                     & \textbf{0.62}                                           & 0.63                     & 0.54                     & \textbf{0.58}                                           & 0.59                     & 0.46                     & 0.52                               & 0.50                     & 0.45                     & 0.47                                                    & 0.23                     & 0.17                     & 0.19                      \\ 
\hdashline
Hover-Net             & 0.82                     & 0.79                     & \textbf{0.80}                                           & 0.58                     & 0.67        & \textbf{0.62}                                           & 0.54                     & 0.60                     & 0.56                                                    & 0.56                     & 0.51                     & \textbf{0.54}                              & 0.52   & 0.47                     & \textbf{0.49}                                           & 0.28                     & 0.35 & \textbf{0.31}            
\end{tabular}
\label{table:detection}
\end{table*}

\bibliographystyle{ieeetr}
\bibliography{references.bib}


\appendix

\setcounter{table}{0}
\renewcommand{\thetable}{A\arabic{table}}

\begin{figure}[b!]
\centering
  \includegraphics[width=0.65\linewidth]{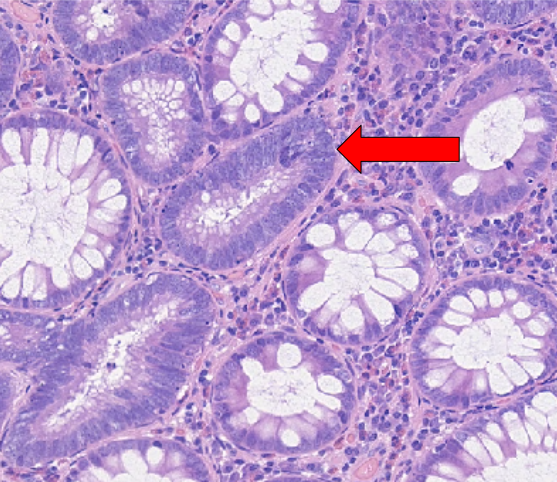}
  \caption{Visual field extracted from colon tissue whole slide image. Red arrow points to a colorectal gland that consists of dysplastic epithelial cells.}
\label{fig:displasia}
\end{figure}

\begin{figure}
  \includegraphics[width=1.0\columnwidth]{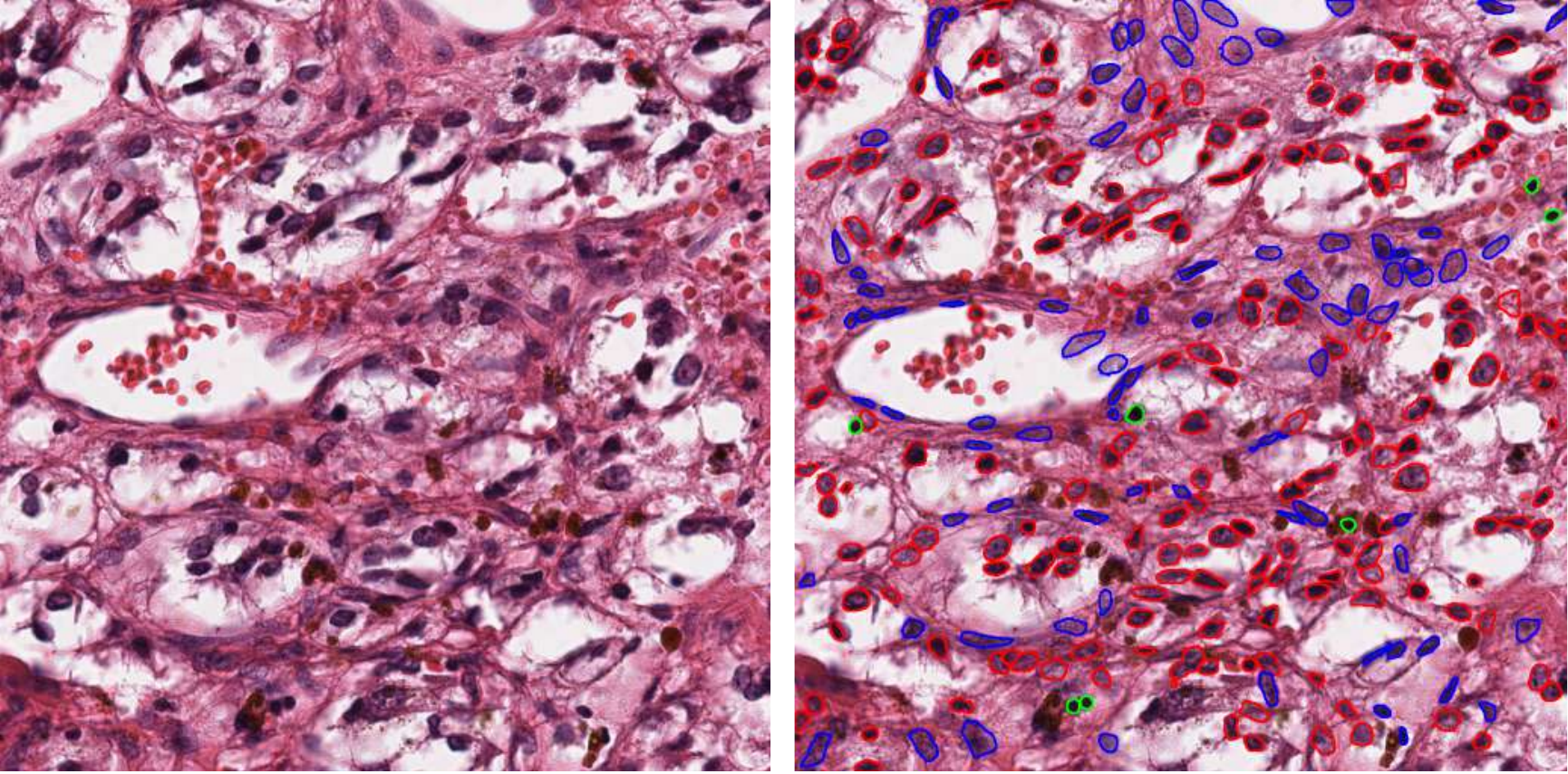}
  \caption{Visual field extracted from adrenal gland tissue and its ground truth on the right.}
\label{fig:adrenal}
\end{figure}

\subsection{PanNuke labelling schema}

The categories in PanNuke consist of: neoplastic, non-neoplastic epithelial, connective tissue, inflammatory and dead cells. These labels can be grouped into either neoplastic or non-neoplastic cell types. However, as Table \ref{table:plasticity} shows, non-neoplastic can cover everything from normal to inflammatory conditions, degenerative, meta-plastic, atypia and dysplasia. In Figure \ref{fig:displasia} we display a visual field from colon tissue where pathologists identified dyspalstic epithelial cells. Here, dysplasia specifically refers to a pre-neoplastic stage, where it has not yet developed into a benign or malignant tumor. Therefore, these colon tissue cells would be labeled as non-neoplastic epithelial. Neoplastic labels in PanNuke specifically correspond to benign and malignant tumor cells. 

\begin{table}[h]
\centering
\caption{Breakdown of neoplastic and non-neoplastic cell types and sub-types.}
\label{table:plasticity}
\begin{tabular}{ll} 
\hline\hline
Plasticity                      & Sub-types         \\ 
\toprule
\multirow{2}{*}{Neoplastic}     & Malignant Tumors  \\
                                & Benign Tumors     \\ 
\midrule
\multirow{8}{*}{Non-neoplastic} & Normal            \\
                                & Hyperplastic     \\
                                & Hypertrophic     \\
                                & Meta-plastic      \\
                                & Inflammatory      \\
                                & Degenerative      \\
                                & Atypia/Dysplasia  \\
\midrule
\end{tabular}
\end{table}

More than 2,000 visual fields from a variety of tissues have been reviewed when developing PanNuke. A major challenge for pathologists during the semi-automatic verification process in PanNuke was the necessity to refer back to the original WSIs to label the categories in the visual field. In Figure \ref{fig:adrenal}, parts of this image can be marked straightforwardly as stroma and inflammatory cells, but classifying the rest of this image is more challenging. For instance, it may represent retraction artifact or neoplasm, specifically pheochromocytoma. Once the WSI was viewed, this area turned out to be neoplastic, i.e. pheochromocytoma. This reiterates the point made in the main text that algorithms are trained on image patches and therefore do not contain the contextual information present in the WSI.

\begin{figure*}
\begin{center}
  \includegraphics[width=1\linewidth]{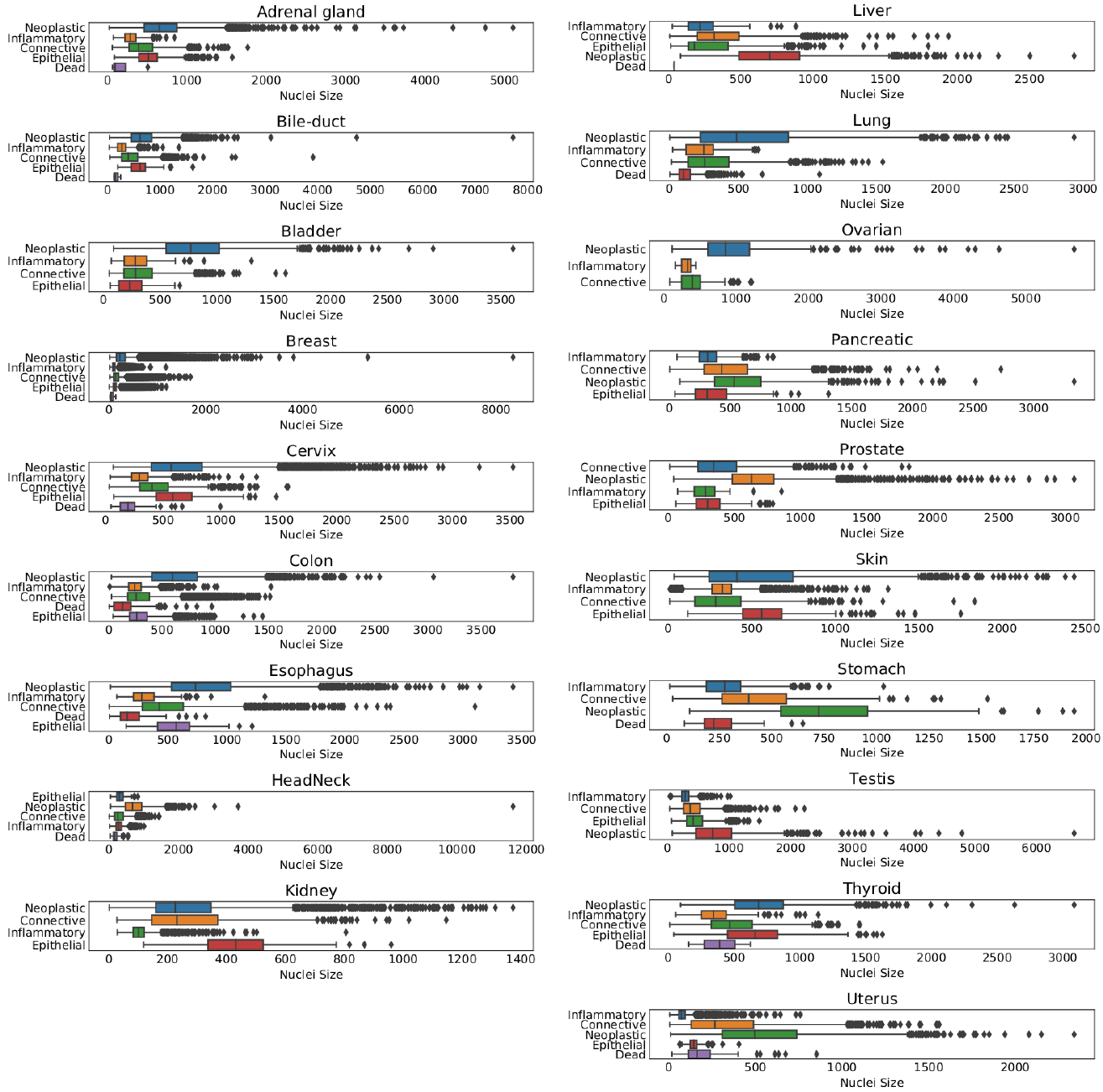}
\end{center}
  \caption{Nuclei size distribution per class within every tissue type in PanNuke. Nuclei size is measured by pixel count within a segmentation mask for a particular nuclei.}
\label{fig:sizes}
\end{figure*}

This is further exacerbated by the distribution of the nuclei size. Nuclei from the connective tissue are noticeably larger - however their shape and size do not directly indicate its category, as presented in Figure \ref{fig:sizes}. This exemplifies the challenge of classifying between non-neoplastic epithelial and connective cell categories. Neoplastic cells tend to be far larger on average compared to any other category. Besides the cell size variability per category within the tissue type, there is a significant difference in distribution of size between the tissues. Epithelial cells in particular vary in size between tissues- this emphasizes the importance of collecting labeled datasets across different tissue types.

\end{document}